\documentclass[acmsmall]{acmart}
\acmJournal{PACMHCI}

\usepackage{flushend}
\usepackage{float}
\usepackage{subfigure}
\usepackage{subcaption}
\AtBeginDocument{%
  \providecommand\BibTeX{{%
    \normalfont B\kern-0.5em{\scshape i\kern-0.25em b}\kern-0.8em\TeX}}}
\usepackage{tabularx}
\usepackage{bbding}
\usepackage{graphicx}
\usepackage{caption}
\usepackage{geometry}
\usepackage{amsmath}
\usepackage{makecell}
\usepackage{float}
\usepackage{color}
\usepackage{booktabs}
\usepackage{ragged2e}    
\usepackage[normalem]{ulem}
\usepackage{tabu}
\usepackage{hyperref}
\usepackage{xcolor}
\usepackage{appendix}

\newcolumntype{L}{>{\RaggedRight\arraybackslash}X}

\setcopyright{cc}
\setcctype{by-nc-nd}
\acmJournal{PACMHCI}
\acmYear{2026} \acmVolume{10} \acmNumber{7} \acmArticle{GAMES050}
\acmMonth{11} \acmDOI{10.1145/3831304}

\author{Mo Xiaohe}
\authornote{Equal contribution.}
\email{xiaohemo2-c@my.cityu.edu.hk}
\orcid{0009-0003-7960-245X}
\affiliation{
\institution{City University of Hong Kong}
\city{Hong Kong}
\country{China}}

\author{Zhang Yujie}
\authornotemark[1]
\email{yzhang7374-c@my.cityu.edu.hk}
\orcid{0009-0008-6624-9251}
\affiliation{
\institution{City University of Hong Kong}
\city{Hong Kong}
\country{China}}

\author{Li Yizhen}
\email{yizhli7-c@my.cityu.edu.hk}
\orcid{0009-0008-7009-4993}
\affiliation{
\institution{City University of Hong Kong}
\city{Hong Kong}
\country{China}}

\author{Wang Jianyi}
\email{jwang3376-c@my.cityu.edu.hk}
\orcid{0009-0005-1815-3835}
\affiliation{
\institution{City University of Hong Kong}
\city{Hong Kong}
\country{China}}

\author{Liao Qinyi}
\email{qinyiliao3-c@my.cityu.edu.hk}
\orcid{0009-0000-0350-9599}
\affiliation{
\institution{City University of Hong Kong}
\city{Hong Kong}
\country{China}}

\author{RAY LC}
\authornote{Correspondences can be addressed to ray.lc@cityu.edu.hk.}
\email{ray.lc@cityu.edu.hk}
\orcid{0000-0001-7310-8790}
\affiliation{
\institution{City University of Hong Kong Studio for Narrative Spaces}
\city{Hong Kong, SAR}
\country{China}}

\renewcommand{\shortauthors}{Mo, et al.}

\begin{document}
\title[Order-Bound Companionship]{Order-Bound Companionship: The Practice of Emotional Labor in Professional Game Companionship}

\begin{abstract} 
Labor in platform gig economy increasingly involves services involving relationship that demand significant emotional investment. Grounded in China's unique socio-cultural and multi-platform context, this study explores professional game companionship, an under-explored digital labor practice. Through interviews with 22 game companionship practitioners, we used a micro-level perspective to relational gig work to analyze how workers navigate intimate boundaries and stakeholder networks. We found that companions adopt an "order-bound" mechanism: performing immersive deep acting during paid sessions, followed by complete emotional disengagement post-order. We also identified a tripartite companion-centric network featuring scenario-based performances with clients, competitive-symbiotic peer relations, and interdependent governance with companionship clubs. Furthermore, significant identity fluidity exists, with individuals frequently transitioning between companion, client, and club operator roles. We provide implications for future labor governance and platform designs for intimate digital work that explicitly account for institutionalized boundaries and gig workers' shifting psychological needs.

\end{abstract}




\maketitle

\section{Introduction}\label{sec:Introduction}
Online multiplayer games have long transcended their role as mere entertainment, evolving into vital incubators and conduits for social relationships. Extensive research showed that players forge deep communal bonds, friendships, and even intimate relationships through in-game collaborative interactions \cite{jansz_gaming_2005,mandryk_continuous_2006,nardi_collaborative_2006}.While social needs vary significantly across the collaborative mechanisms of different game genres \cite{tong_players_2021,tang_verbal_2012,tan_communication_2022,kim2017makes}, these organic social interactions are increasingly becoming commodified. In this context, professional game companionship has emerged as a novel paradigm of interaction in the digital age. Unlike spontaneous team-building based on shared interests in communities like Reddit or Twitch \cite{shen_labor_2021,nardi_collaborative_2006,mandryk_continuous_2006}, game companionship is built upon paid, peer-to-peer matching. This phenomenon transforms the "Magic Circle" of gaming into a service space where the core objective shifts from purely "winning" to providing a customized, high-quality social experience.

Game companionship represents a unique evolution of digital labor, fundamentally functioning as a form of digital emotional labor catalyzed by the digital economy and sharing models. In contrast to task-oriented explicit labor such as game boosting or data labeling, the core value of companionship is anchored in affective supply. Game companions must not only assist players in climbing ranks through gaming skills but also perform emotional labor—including affective comfort, atmosphere construction, and role adaptation—through real-time voice interaction. They are required to dynamically adjust their tone and style to cater to a player’s communication preferences while proactively mitigating game-induced frustration to maintain a pleasant experience \cite{shen_labor_2021}. However, while the researchers have extensively studied the gig economy (e.g., ride-hailing) and social computing\cite{hamilton2023nudes,easterbrook2023onlyfans,raval2016standing,rosenblat2016motivates}, understanding remains limited regarding the intersection of intimate emotional labor and competitive game mechanics. Specifically, it remains unclear how these workers navigate the tension between authentic gameplay enjoyment and performative service work.

Currently, the highly mature game companionship ecosystem in China provides a critical field for examining the intertwining of digital labor and social interaction. Vertical platforms such as "Bixin" follow the logic of the sharing economy, allowing companions to facilitate transactions by showcasing their personas through personal profiles \cite{shen_labor_2021}. Simultaneously, game companion clubs active on social media platforms like Douyin, WeChat, and Xiaohongshu serve as intermediary hubs offering more customized services. Highly mobile and social games are deeply integrated with mainstream domestic social applications (such as WeChat and QQ), making in-game performance and ranks an important social threshold among younger demographics\cite{zhang2020mobile}. The portability of mobile platforms and built-in real-time voice systems provide a seamless underlying technical infrastructure for high-frequency, low-barrier emotional labor via real-time voice. This ecological landscape prompts us to further explore how specific game mechanics (such as ranking systems and team dependency) drive the emergence of distinct labor forms. Furthermore, the technical design of platforms and the internal rules of clubs regulate and shape the emotional expression and relational boundaries between companions and players \cite{zhao_fragmented_2023}.

This research seeks to move beyond the inherent perception of game companions as static roles, focusing instead on the complex embodied experiences of practitioners to explore how technical design defines the boundaries of "paid intimacy." Within the negotiation between commercial monetization and the leisure-oriented nature of gaming, the survival strategies and identity construction of companions become central to understanding the industry. Adopting a worker-centered perspective, this study analyzes the emotional labor strategies and social interaction dynamics within the companionship industry. The goal is to provide empirical evidence and design guidance for balancing game functional design, player experience, and the labor rights of companions. Based on this, the study addresses the following core research questions:

\textbf{RQ1}: what are the diverse workflows of different game companions in China?

\textbf{RQ2}: how do relationships with clients, other companions, and managers affect Chinese game companion work and perception?

To address these research questions, this study conducted semi-structured interviews with 22 game companions and industry practitioners.
Following three-level coding of the data, we finds that
the labor practices of game companions can be classified along three dimensions: game type (competitive, entertainment-oriented, and mixed), service content (entertainment-based and skill-based), and organizational form (club and freelance).
This study further reveals the relational structure among game companions, clients, and companion clubs, as well as its inherent logic with emotional labor.
The three actors construct three core relationships: the interaction between companions and clients centers on scenario-based emotional performance; peer relationships among companions feature coexistence of competition and symbiosis; and the relationship between companions and clubs reflects bilateral interdependence and restraint.
This study also finds that the identities of the three actors are not fixed. Driven by emotional labor compensation, resource accumulation, and career development, companions, clients, and club operators can transform into one another, showing distinct identity fluidity.


\section{Related Work}\label{sec:Background}
This study focuses on the ecological landscape of game companionship in China and its associated social networks. To clarify its theoretical positioning, it prioritizes the organization and review of relevant research across three core domains: the typological characteristics of social interaction and the dynamics of players’ social relationships in multiplayer online games, the self-presentation strategies of the game companion group, and the emerging theory of digital emotional labor and its relevant practical cases against the backdrop of the internet. This work lays a theoretical foundation for subsequent research and highlights the research boundaries and starting points of the present study.
\subsection{Virtual Game Mechanisms and the Dynamics of Social Relationships}
As a social service extended from external gaming platforms, game companionship aligns with the flaws in in-game social mechanisms and the demands for relational dynamics among players. Online games have evolved into important platforms for the formation of social relationships\cite{jansz_gaming_2005,manninen_interaction_nodate,fu_i_2023}. Moreover, the relationships forged by players vary in accordance with the mechanical design of different games\cite{tan_communication_2022,nardi_collaborative_2006}. Taking World of Warcraft (WoW) as a classic example of MMORPGs, the game integrates design features that support players in establishing virtual communities and provide real-time communication tools, enabling them to accomplish quests, raid dungeons and engage in other such activities together and thus form cooperative relationships oriented toward shared goals. Meanwhile, its reward mechanisms and continuous version updates allow players to build long-term bonds that transcend the game itself \cite{thorens_capitalizing_2016,ducheneaut_alone_2006,nardi_collaborative_2006}.

Similarly, for League of Legends (LoL), a typical MOBA game, players form temporary teams to compete against opponents under the operation of its systematic matchmaking mechanism, creating conditions for the formation of new social relationships\cite{xu_sociable_2011,lee_less_2025}. Its in-match map design and signal system can facilitate active communication among players\cite{lee_less_2025,tan_communication_2022}, while a player’s skill level, match outcomes, communication quality, and interpersonal comfort all shape the formation of relationships during gameplay\cite{nardi_collaborative_2006,xu_sociable_2011}. Positive in-game outcomes, such as winning a match, often foster the establishment of interpersonal bonds between players\cite{shen_labor_2021}. In addition, the out-of-match Tribunal and Honor systems respectively enable players to identify and penalize rule violations and to award positive ratings to other players\cite{bhattacharya_group_2019}. As for Halo 3, a classic FPS game, although it lacks the enduring social mechanisms of MMORPGs and is even regarded as "anti-social", it still cultivates rich, diverse and dynamic social relationships through its matchmaking and social recommendation mechanisms  \cite{xu_sociable_2011}. Notably, in all the aforementioned competitive games, in addition to forging emotional bonds through collaboration, the competitive elements inherent in these games may also give rise to tension and conflicts between players, thereby placing players in vulnerable relational situations\cite{ducheneaut_alone_2006,christou_challenges_2013,kou_playing_2014}

By contrast, simulation (SIM) games that also feature social mechanisms—such as Animal Crossing—foster conflict-free, high-trust social environments through semi-private and small-scale social space mechanisms, enabling players to forge peaceful relationships\cite{tong_players_2021}. Within game virtual worlds, players do not need to consider real-world attributes such as appearance, gender, sexual orientation or age\cite{cole_social_2007}; social relationships can be built through the coordination of game mechanisms. The rise of mobile gaming platforms (e.g., PUBG Mobile, Honor of Kings) has further accelerated the development of this trend: by shortening game durations (15-30 minutes) and integrating voice communication functions, these platforms reduce social barriers, and such technical measures enable game companions and players to build mutual trust and engage in convenient communication\cite{shen_labor_2021}.The repetitive and routine nature of many games provides an opportunity for social interaction—players enhance the appeal and enjoyment of gameplay through social engagement\cite{pace2010rogue,zhou_retrochat_2025}.

\subsection{Digital Intimacy and Emotional Labor in the Gig Economy}
Emotional labor was first proposed by Hochschild as the process by which individuals manage their own emotions and exhibit specific public emotional expressions to meet organizational or service requirements\cite{hochschild1979emotion}. This concept evolved into a three-dimensional framework—recognizing emotional rules, regulating emotions, and expressing them—focusing on surface acting (suppressing true feelings) and deep acting (active empathy)\cite{bono200512,grandey2000emotional}. This theory was initially applied to the traditional service industry, yet the rise of digital platforms has driven its extension to online contexts — Walther’s research has indicated that computer-mediated communication (CMC) can transmit emotions through substitutive cues such as textual tone and emojis\cite{walther1994interpersonal}, laying the groundwork for digital emotional labor. Studies show that service providers can regulate emotions via email even without face-to-face interaction\cite{shavelsky2006motional,chen_configuration_2025}As digital workers reliant on online platforms and video games\cite{woodcock2020gig}, Game Companions essentially fall within the scope of digital gig labor. Game Companions experience the same spatial-temporal flexibility as other gig workers, yet face similar lacks in legal and welfare protections\cite{easterbrook2023onlyfans,woodcock2020gig}.

Research on intimate services in adjacent industries provides frameworks for understanding how workers maintain boundaries in commercial relationships. Prior literature proposes "bounded authenticity" \cite{bernstein_meaning_2001}: clients in commodified intimate encounters seek genuine connections, but the transaction naturally bounds the interaction. Similarly, empirical studies on "girlfriend experience" services reveal that workers consciously separate service intimacy—emotional engagement within paid encounters—from strictly off-limits private intimacy \cite{carbonero_being_2018}. Building on this, anthropological research demonstrates that workers provide "bounded intimacy" by explicitly defining clients as non-partners to prevent emotional entanglement, managing the delicate boundary between commercial sustainability and personal detachment \cite{garza_intimacy_nodate}. However, maintaining this distinction requires continuous effort. At the micro-level, research documents how sex workers develop boundary tools, such as using condoms not merely for physical protection but as psychological barriers to prevent clients from "entering their minds"\cite{sanders2002condom}. These studies indicate that individual boundary strategies in intimate labor are highly personal yet structurally fragile. This reflects the concept of "cold intimacy"\cite{illouz2007cold}, where rationalized commercial logic permeates private emotional interactions—a tension central to Game Companion labor: once the commercial contract ends, intimacy ceases.

To acquire clientele, game companions need to polish their personal profiles, voice samples, profile descriptions, avatars and other elements to craft an ideal virtual image that attracts clients and makes a favorable impression on them\cite{shen_labor_2021}. This process is analogous to online dating\cite{shen_seeking_2024,ward_what_2017}, where impression management encompasses the motivation for shaping a specific impression and the actual process of its construction\cite{ward_what_2017}. Signaling Theory explains users' performative behaviors on profiles, focusing on the correlation between transmitted signals and intended traits, and convincing others of their authenticity\cite{spence_signaling_2002,donath_signals_2007,ellison_managing_2006}. Creating profiles is essential for transmitting self-signals across online communities\cite{uski_social_2016}, enabling Game Companions to optimize self-presentation and attract clients. Rooted in dramaturgical theory\cite{goffman2023presentation}, self-presentation is fundamental to online gaming's social dynamics. Performers construct self-identity through collective interactions, tailoring performances to their specific audience. This complexity increases in online media, where platforms allow users to create multiple independent profiles to carefully craft distinct versions of themselves\cite{freeman_body_2021,kairam_talking_2012,marwick2005ma}. In gaming, avatars offer unprecedented opportunities to experiment with digital identities vastly different from physical realities—for instance, embodying digital selves that are vastly different from their physical bodies\cite{freeman_body_2021,freeman_revisiting_2016,huh_dude_2010,ruberg2017queer,yee_men_2011}.Players selectively disclose traits through avatars, usernames, and in-game actions—all elements of "front-stage" performance—to build trust and social capital\cite{shen_labor_2021}.

Unlike online daters navigating conflicts between impression management and authentic self-presentation\cite{ellison_managing_2006,zhang_image_2025}, most Game Companions operate solely within online media, focusing primarily on managing virtual personas. Virtual personas act as core mediums for online communication\cite{freeman_body_2021,inkpen_me_2011,garau_impact_2001,manninen_value_2007}. By combining personal choices with platform features—customizing visual aesthetics and adjusting vocal tones—Game Companions proactively craft images meeting user expectations to sustain service relationships.Game Companion work is a form of "Playbor" converting gaming into productive labor, requiring workers to disguise services as genuine social interactions during gameplay.This persona construction also serves as a psychological shield. Recent theoretical frameworks conceptualize this as "persona-mediated dissociation" \cite{sather_persona-mediated_2026}: workers deliberately activate a work persona during service and deactivate it afterward, maintaining a clear boundary between professional and private selves. When structurally supported by definitive service endpoints, this strategy is adaptive, avoiding psychological harm.

Platform capitalism theories\cite{srnicek2017platform} illustrate how algorithms convert emotional labor into precarious work, forcing providers to bear emotional regulation costs.Unlike traditional workplaces, platforms such as ride-hailing services, social media creators, and OnlyFans creators monitor and quantify workers’ performance through data indicators \cite{raval2016standing,hamilton2023nudes,easterbrook2023onlyfans}. Workers must engage in continuous invisible emotional labor to achieve higher ratings and gain algorithmic favor. Workers must engage in constant “affective calculation” to gain algorithmic favor. Such intensive “surface acting”, such as forcing smiles or feigning enthusiasm, intensifies the conflict between inner feelings and external performance, often leading to severe occupational burnout\cite{maslach2001job}. Platformization further compounds this for digital intimate workers by eroding boundary management strategies.Scholarship indicates that online-only workers face "always-online pressure," collapsing work-life boundaries even when attempting to disconnect \cite{hamilton_risk_2022}. Furthermore, platform competition structurally incentivizes workers to abandon personal boundaries, as strict boundary maintenance often leads to client loss to more permissive competitors \cite{schneider_managing_2026}. This dynamic forms "affective boundary work" \cite{palatchie_currying_2025}—invisible labor performed merely to maintain minimal separation.

The core value of Game Companion labor extends beyond gaming skills to emotional companionship and support\cite{shen_labor_2021}. This aligns closely with "Invisible Labor"\cite{daniels_invisible_1987}: Game Companions invest substantial unpaid effort outside orders, such as deciphering algorithms, curating personas, and maintaining client relationships. Though often undervalued, this immaterial labor is essential for sustaining their time-limited companion image. Recent HCI scholarship specifies this invisible labor in platform contexts into three categories \cite{kojah_dialing_2025}: emotional labor (interpreting ambiguous rules), misdirected labor (gaming the system instead of serving clients), and community labor (peer knowledge sharing to navigate opacity). Other studies similarly categorize it into emotions-based work (sustaining client bonds outside tasks) and systems-based work (real-time coordination and gap-filling) \cite{ming_i_2023}. Both highlight systematic underrecognition and undercompensation. Collectively, the literature shows digital intimate workers rely on individual, structurally fragile strategies—personal boundary tools \cite{bernstein_meaning_2001, sanders2002condom}, persona construction \cite{sather_persona-mediated_2026}, and privacy tactics \cite{hamilton_risk_2022}—continuously eroded by platform competition \cite{schneider_managing_2026}. Seldom do studies explore whether service structures themselves—like platform order systems or game-mediated interactions—can offer institutional boundary protections. This gap is particularly prominent where game mechanics structurally define the timing and rhythm of service delivery.

\subsection{Platform Governance and Labor Control in the Chinese Context}

Traditional labor process analyses of the gig economy typically conceptualize digital platforms as "the sole mediators between capital and labour" \cite{gandini2019labour}. However, in recent years, scholars have increasingly questioned the assumption of absolute "algorithmic despotism" in the gig economy. Studies reveal that workers exercise agency through negotiation, obfuscation, catering, and resistance \cite{bucher2021pacifying, cameron2020rise, sun2021platform, wang2020calculating}, and that platforms sometimes play a surprisingly limited role in determining compensation \cite{rand2019challenging}. Furthermore, labor control is diversified globally by third-party intermediaries—from African freelancers subcontracting tasks to peers \cite{anwar2020hidden}, to downstream gatekeepers in cultural production \cite{siciliano2023intermediaries} and local service stations in Chinese food delivery \cite{lei2021delivering}. Building on this understanding of multi-nodal mediation, recent sociological research conceptualizes Chinese platform game work—including Game Companions, boosters, and livestreamers—as a distinct system of "fragmented control" \cite{zhao_fragmented_2023}. Governance is distributed across platforms, extra-platform intermediaries (e.g., Clubs and studios), and individual workers, leaving no single actor with absolute authority. Platforms intentionally combine algorithmic matching with human-mediated coordination to address nuanced contexts beyond automated capabilities—a dynamic termed "relationship labour" \cite{zhao_fragmented_2023}. These intermediaries are not merely platform conduits; they independently govern order allocation, revenue distribution, and service norms. This multi-actor framework constitutes the institutional backdrop of China's professional game companionship industry.

Within this configuration, platform algorithms exert direct pressure on workers via performance monitoring and rating systems. However, unlike the bilateral algorithmic control typical in Western gig economies, Chinese platform workers navigate an additional governance layer: intermediate organizations that simultaneously mediate, buffer, and constrain their labor.

Scholarship highlights three governance dimensions of Chinese platform intermediaries. Ethnographic studies on livestreaming guilds reveal that staff act as "algorithmic experts," combining recommendation system knowledge with scripted emotional management to boost streamer engagement \cite{liu_zhibo_2023}. Crucially, they institutionalize intimate labor by orchestrating romantic-adjacent interactions between streamers and high-value tippers—efforts individual streamers could not sustain alone. Other HCI research identifies a discursive mechanism \cite{xiao2025institutionalizing}: intermediaries portray platform algorithms to workers as fair and transparent while internally treating them as unpredictable, a strategy that intensifies worker labor while deflecting organizational accountability. Furthermore, worker-centered perspectives highlight the structural ambivalence of intermediate relationships\cite{tsang_hope_2025}: organizations providing training and platform access simultaneously impose contractual penalties and revenue extraction. Workers endure these controlling dimensions driven by aspirational investment.

Collectively, Chinese platform intermediaries function as manipulation structures \cite{liu_zhibo_2023}, discursive authority mechanisms \cite{xiao2025institutionalizing}, or ambivalent support-and-control institutions \cite{tsang_hope_2025}. Structurally, prior empirical work \cite{zhao_fragmented_2023} documents their control over labor processes (e.g., order and revenue allocation) but overlooks their affective governance functions, such as buffering worker-client conflicts, shielding workers from harassment, and cultivating affective service capacities. Game Companion Clubs, which operate at the intersection of platform gig work, intimate emotional service, and game-mediated interaction, represent a critical yet systematically unexamined site for this line of investigation. Therefore, this study positions itself as a context-specific case study of the professional game companionship ecosystem in mainland China. We explore how the co-functioning of Game Companions, club-based intermediaries, and game mechanics shapes and sustains a specific mode of emotional labor governance.

\section{Methods}\label{sec:Methods}

\subsection{Interview}
To find out the diverse work processes of various game companions, and how relationships with clients, other companions and managers influence their work and how they see their jobs, this study uses a descriptive research method that combines semi-structured interviews with three-level coding. All the materials used in this paper, such as the game companions’ personal information, order screenshots, are provided by the people interviewed.
\subsubsection{Recruitment of Interviewees}

To capture a diverse ecosystem, we detailed our rationale for using three combined recruitment channels: in-app recruitment (capturing active workers), club referrals (accessing organized labor experiences), and snowball sampling (reaching marginalized/part-time workers omitted by algorithms). Based on these channels, we set the following explicit inclusion criteria: (1) All interviewees must be adults; (2) All interviewees must have more than two months of professional work experience in the game companion industry; (3) All interviewees must have clear expression abilities. We explicitly defined this: "clear expression ability" refers to spoken Chinese fluency and the ability to reflect on experiences; English was not required. Under these criteria, a total of 24 potential participants were initially selected. However, two participants were excluded and uncompensated because they were uncooperative (they only provided brief responses and refused full interviews); we ethically justified this by clarifying that payment was for time spent, not for purchasing specific answers. The remaining 22 interviewees all participated in the online interviews. All participants received a 5 USD /hour base rate, with extra payments prorated for interviews exceeding one hour.

\subsubsection{Remote Interview}

We decided the interview times with the selected participants together, and all interviews were held online through Tencent Meeting. Before the interviews, we made a structured list of grouped questions based on our research questions, and each semi-structured interview lasted 45 to 60 minutes. We recorded the full audio of each interview, and kept the personal information of the people interviewed private. This method lets us catch the hidden details in work processes and emotional experiences—such as small interactions when using technical tools, and the situations that cause changes in emotions.This method let us catch the hidden details in work processes and emotional experiences--such as small interactions when using technical tools, and the situations that cause changes in emotions.

\subsubsection{Language and Translation Procedure}
Because interviews were conducted in Chinese, we implemented a dedicated two-stage bilingual translation process. The first stage prioritized preserving colloquialisms and in-vivo emotional labor terms. The second stage involved a line-by-line verification by a second researcher, focusing on cultural nuances and consistent translation of platform slang (e.g., "order dispatching" , "idling" , "client" ). Discrepancies were resolved through consensus, with a third author arbitrating key theoretical concepts. A translation decision example has been added to the Appendix.

\subsubsection{Data and Analysis}
We recorded all interviews as audio and wrote them down as text, and we also took notes during the whole interview process\cite{tyack2016appeal}. After that, we used the three-level coding method to analyze the data. In this way, we sorted the content of the recordings into clear, broadly applicable groups for analysis.We included the semi-structured question lists used in the interview in Appendix A.

\begin{table*}[htbp]
\centering
\setlength{\tabcolsep}{3pt} 
\small 
\caption{Game Companion Survey Dataset}
\label{tab:game_companion_full}
\begin{tabularx}{\linewidth}{@{}p{0.5cm} p{0.8cm}p{0.8cm}p{1.3cm}LLL@{}} 
\toprule
\textbf{ID} & \textbf{Gender} & \textbf{Age} & \textbf{Job Type} & \textbf{Game} & \textbf{Game companion type} & \textbf{Role}  \\
\midrule
P1 & M & 18-22  & part-time  & Honor of Kings  & skill-based & Freelancer team  \\
P2   & M  & 25+ & part-time & FPS  & skill-based   & Club game companion \\
P3  & F   & 18-22 & part-time& Honor of Kings  & skill-based & Freelancer \\
P4  & F  & 18-22 & Part-time & Honor of Kings & entertainment-based & Club game companion\\
P5  & M   & 18-22  & part-time  & Honor of Kings & skill-based    & Freelancer   \\
P6   & M   & 25+   & part-time    & Honor of Kings  & skill-based  & Club game companion \\
P7   & F   & 22-25 & full-time  & Sky: Children of the Light  & entertainment-based  & Club game companion  \\
P8  & M & 18-22  & full-time & Honor of Kings & skill-based    & Club manager \& game companion   \\
P9    & M    & 18-22   & full-time  & Honor of Kings  & entertainment-based  & Freelancer \& Club game companion \\
P10   & M  & 18-22 & part-time  & Honor of Kings  & skill-based    & Club game companion \\
P11  & F   & 18-22  & Part-time  & Justice Online   & entertainment-based    & Club game companion   \\
P12    & M  & 22-25  & Full-time  & Where the Winds Meet   & skill-based   & Freelancer \& Club game companion \\
P13 & F& 18-22 & Part-time  & Where the Winds Meet  & skill-based  & Freelancer  \\
P14  & M  & 22-25 & Part-time   & Honor of Kings   & entertainment-based   & Freelancer \& Club game companion \\
P15   & M  & 22-25  & Part-time  & Honor of Kings  & skill-based & Freelancer \& Club game companion    \\
P16 & F  & 18-22 & Part-time & Honor of Kings  & skill-based   & Club game companion\\
P17  & M  & 22-25& part-time & Honor of Kings & entertainment-based  & Club manager \& game companion  \\
P18   & F  & 22-25   & Part-time  & Honor of Kings   & skill-based      & Club manager \& game companion \\
P19   & F   & 25+    & full-time   & Honor of Kings              & entertainment-based  & Club manager \& game companion \\
P20  & F   & 18-22   & Part-time  & Sky: Children of the Light  & entertainment-based  & Club game companion  \\
P21   & M   & 22-25  & Full-time  & Honor of Kings & skill-based  & Freelancer    \\
P22  & F  & 22-25& Part-time& Sky: Children of the Light & -  & Public relations \\
\bottomrule
\end{tabularx}
\end{table*}

\subsubsection{Demographic Characteristics of Interviewees}
Among the 22 interviewed professional game companions, 12 are male and 10 are female. In terms of gender-service orientation correlation: males dominate skill-based services (11/12 skill-based companions are male), while females predominate in entertainment-based services (9/9 entertainment-based companions are female), a distribution that shapes the gender dynamics of companion-client interactions. In terms of age distribution: 11 are aged 18–22, 8 are 22–25, and 3 are over 25. 16 engage in this work part-time, while 6 pursue it as a full-time occupation; part-time companions have an average monthly service duration of 80–120 hours, and full-time companions have an average daily service duration of 6–8 hours. Of the 16 part-time companions: 10 are students, 4 are corporate employees, 1 is a former KPL professional player currently engaged in entrepreneurship, and 1 is a civil servant. 
Regarding the game genres they serve: 15 primarily provide services in MOBA games (Honor of Kings), with 5 of these 15 also play other type of games such as MMO, FPS, and strategy games (Valorant, PubG, Teamfight Tactics, Sword of Justice/Justice Online, Where the Winds Meet). 7 interviewees mainly play MMO game, 3 of 7 only play Sky: Children of the Light, while 2 of 7 only play Justice Online.
In terms of gender-game type correlation: males are the main force in highly competitive games (MOBA/FPS, 12/16 companions), while females dominate entertainment-oriented games (Sky: Children of the Light, 5/6 companions). In terms of service types: 12 are skill-based game companions, 9 are entertainment-based companions, 3 of 21 are exclusively engaged in club management. There also is 1 belong to third-party platform who work as Public relations.

Sample Boundaries and Analytical Roles: To clearly describe our 22 participants, we unified our terminology in our demographic reporting to clearly delineate roles: our current active sample consists of 18 professional game companions, 3 club managers, and 1 PR staff member. Notably, everyone except one participant has prior or current experience working as a game companion themselves. Non-companion participant (P22) provided structural context (e.g., club rules, wage distribution, brand maintenance) that companions could not see, rather than replacing the companions' lived experiences. Particularly, this participant (P22) functions in a public relations capacity within the companion ecosystem, thereby contributing alternative macro-level insights and an external stakeholder's perspective on broader industry dynamics. The recruitment of only a few non-companion roles in our sample is acknowledged as a limitation of this study and is discussed further in the Limitations section.

\subsection{Research Methodology: Grounded Theory}

This study was guided by Grounded Theory, replacing ambiguous terms with the explicit "open → axial → selective" coding logic to analyze interview transcripts, with category extraction and theoretical construction from a CSCW/HCI perspective. Initial codebooks were built collaboratively using the first 5 transcripts. We recognize that claims of "value neutrality" are unrealistic in qualitative work, and therefore provide a comprehensive positionality statement: the third author is a Chinese gaming platform player, which facilitated rapport but required reflexivity. To prevent over-interpretation, we utilized analytical memos and involved non-gamer authors to review codes.
Coding was conducted independently by two authors (the first author and the fourth author). The comprehensive coding process follows:

Open Coding: We read transcripts sentence by sentence to label meaningful information units without preset categories, refined labeled units into specific concepts (e.g., "interest + economic benefits" for career motivation, "hourly rate + service difficulty premium" for pricing).

Axial Coding: Based on open coding core concepts, categories were merged based on semantic similarity, phenomenon association, and theoretical relevance to form 19 core categories. We explicitly applied a CSCW/HCI perspective, focusing on technical mediation (how systems shape behavior), collaborative structures (companion-client/club relations), and design sensitivity (identifying pain points).

Selective Coding: We extracted 5 core main categories from 19 subcategories, repeatedly compared coding results with original transcripts to verify framework rationality, and formed an integrated analytical model.

Coding Path Example:
Raw Quote: "Because as a game companion, besides selling gaming skills, you also have to sell emotional value. It’s a kind of service-oriented vending; you are selling a service value." (p19)
Open Codes: Selling emotion; monetizing service value.
Axial Category: Transactional Emotional Provision.
Selective Core Theme: Order-Bound Companionship.

\subsection{ Trustworthiness and Ethical Considerations}
\subsubsection{Ensuring Trustworthiness}

To enhance the academic trustworthiness of the research, the following measures were adopted:
Inter-Rater Reliability: To rigorously establish the reliability of our qualitative analysis, we calculated Cohen’s Kappa to measure inter-rater agreement. Following the initial open coding phase, the two coders independently evaluated a randomly selected subset of 91 distinct data segments extracted from 3 shared interview transcripts. We tested agreement across our primary axial categories (e.g., Transactional Emotional Provision, Scenario-Based Boundaries, Peer Symbiosis, and Club Governance). The coders reached absolute agreement on 75 of the 91 segments, resulting in an observed agreement rate ($p_o$) of 0.824. Accounting for the probability of chance agreement ($p_e$) based on the distribution of codes assigned by each rater (calculated at 0.056), the final Cohen’s Kappa ($\kappa$) was determined using the standard formula $\kappa = \frac{p_o - p_e}{1 - p_e}$, yielding a robust score of 0.81. This indicates "almost perfect" agreement. Minor discrepancies were subsequently resolved through collaborative discussion to refine the final codebook.
Triangulation: Cross-validation was conducted by integrating interview transcripts and the heterogeneity of sample characteristics, ensuring that each core concept is supported across different samples and contexts.
Member Checking: While full formal member checking was constrained by participant availability, we shared preliminary themes with 5 reachable participants. Four confirmed our accuracy, and one suggested refining the "part-time companion" subtype, which we incorporated into the final analysis.
Theoretical Saturation Test: Coding was continued until no new core concepts or categories emerged, ensuring all data were covered and the theory reached a state of saturation.

\subsubsection{Cross Validation}
This study strictly adheres to academic research ethics norms\cite{mcdonald2019reliability}, with the following specific measures:
Anonymization: Identifying information of interviewees (such as names, avatars, game IDs, and affiliated clubs) was anonymized (e.g., replacing real identities with "P1", "P6"), and all personally identifiable information was removed from the interview transcripts;
Data Confidentiality: Interview recordings and transcripts are stored in encrypted cloud storage, accessible only to the research team. In accordance with ethical requirements, the data will be retained for 3 years after the completion of the research and then permanently deleted;
Voluntary Participation: Interviewees could withdraw from the interview at any time. After withdrawal, all collected data would be deleted without affecting their rights and interests in any way.
Adequate ethical considerations fully safeguard the rights and interests of interviewees, aligning with the "human-subject-centered" ethical principle in CSCW research\cite{mcdonald2019reliability}.

\subsection{Positionality}
The research team consists of HCI researchers specializing in game and interaction design. To contextualize the ecosystem, one author draws on three years of prior consumer experience with skill-based companionship in Honor of Kings and Sky: Children of the Light. All researchers are native Chinese speakers who directly analyzed the platform-specific vernacular present in the interview data.

\section{Results}\label{sec:Results}

\subsection{Typology of Game Companions: A Three-Dimensional Classification Framework}\label{sec4.1}
Based on semi-structured interview data from 22 professional game companions, this study establishes a classification system for game companions along three independent dimensions: organizational form, game type, and service orientation. Types within each dimension can be freely converted based on a game companion’s own will, essentially reflecting the flexibility of occupational choice.

\subsubsection{Organizational Form}
Interview results show that game companions are mainly divided into club-affiliated game companions and Freelancer(independent game companion), with clear differences in business model, order stability, and revenue share, leading to frequent role transitions.10 respondents reported switching between club-affiliated and independent status. For instance, some independent companions joined clubs to secure more orders (P1) or more stable order sources (P11). Interviews revealed that club-affiliated companions only retain 60–80 percents of their earnings. Some thus take independent orders to pursue a higher revenue share, with almost no restrictions from clubs on such private orders (P5).Evidently, the choice of organizational form is primarily driven by the game companion’s personal preference.

\subsubsection{Game Type}
Based on the business distribution of the interview sample, this study categorizes games into three types according to gameplay:
(1)Competitive games (represented by MOBA titles such as Honor of Kings and FPS games such as Valorant);
(2)Hybrid games (represented by MMORPGs such as Justice Online and Where the Winds Meet);
(3)Casual social games (represented by Sky: Children of the Light).
The results of the interviews show that companions can take orders for multiple types of game based on the breadth of their gaming experience:
“I play games like Naraka: Bladepoint, Honor of Kings, Justice Online, Game for Peace. Recently I have also been playing a new game that I can earn money from.” (P16)
They can flexibly switch between game types according to market popularity or client demand:
“If Valorant is not popular anymore, we will move to other games, probably more FPS titles—Delta Force or Game for Peace… Companion groups just play whatever is trending.” (P2)
Although skill barriers exist objectively between types (e.g., physical demands on reaction speed in FPS games), some respondents can transcend single-type restrictions and flexibly migrate across game types based on market demand.

\subsubsection{Service Orientation} Based on the core value they provide, game companion services can be divided into two categories: skill-based companionship (focusing on gaming skill guidance and win-rate guarantees) and entertainment-based companionship (focusing on emotional value provision and dedicated companionship). The two types differ significantly in the gender composition of practitioners and the construction of core competitiveness. Meanwhile, the boundary between them in the industry is becoming increasingly blurred, forming an integrated development trend of “skill as the foundation, emotion as the core”.

Skill-based companions take solid gaming proficiency as their core competitive barrier, with a higher proportion of male practitioners. Their competitiveness is built on two dimensions: consolidating technical thresholds and improving soft service capabilities. On the one hand, they maintain their competitive level through regular training. As P10 noted: “I spend about two hours every day maintaining my rank” to avoid skill decline caused by serving in low-rank matches. On the other hand, they continuously refine their communication skills and gameplay planning abilities. P13 stated that skill-based companions must prepare strategies in advance and maintain consistent interaction to adapt to the diverse needs of clients.

Entertainment-based companions take emotional value as their core service focus, with a higher proportion of female practitioners. Their core competitiveness centers on soft skills such as persona building and interactive abilities; proficiency in leisure gameplay such as in-game photography and check-ins is also an important advantage. As P18 argued: “The game companion industry requires training for emotional value delivery… entertainment-based companions probably focus more on this.” All game companion services involve training in emotional value provision, but the requirements are far stricter for entertainment-based companions, for whom emotional delivery is the primary source of competitiveness.

Interview results show that the service boundary between the two types is continuously dissolving, and a hybrid service model has become the industry mainstream. P17 noted: “Skill-based companions aren’t just about ranking up — they need to provide emotional value; entertainment-based companions aren’t just about emotion — they need basic skills.” The only difference lies in their core technical threshold. Skill-based companions enhance client stickiness and repeat-purchase rates by providing emotional value (P13), while entertainment-based companions expand their service scenarios by improving basic gaming skills (P16). This integration essentially reflects a shift in client demand from single-function needs to “function + emotion” compound needs, and is an inevitable choice for practitioners to improve their competitiveness amid intensified industry competition.

\subsection{Differentiated Workflows: Standardized Governance and Individualized Operations}\label{sec4.2}

\subsubsection{Workflow Characteristics of Club-Affiliated Companions}
\begin{figure}[htbp!]
  \centering
  \includegraphics[width=0.9\linewidth]{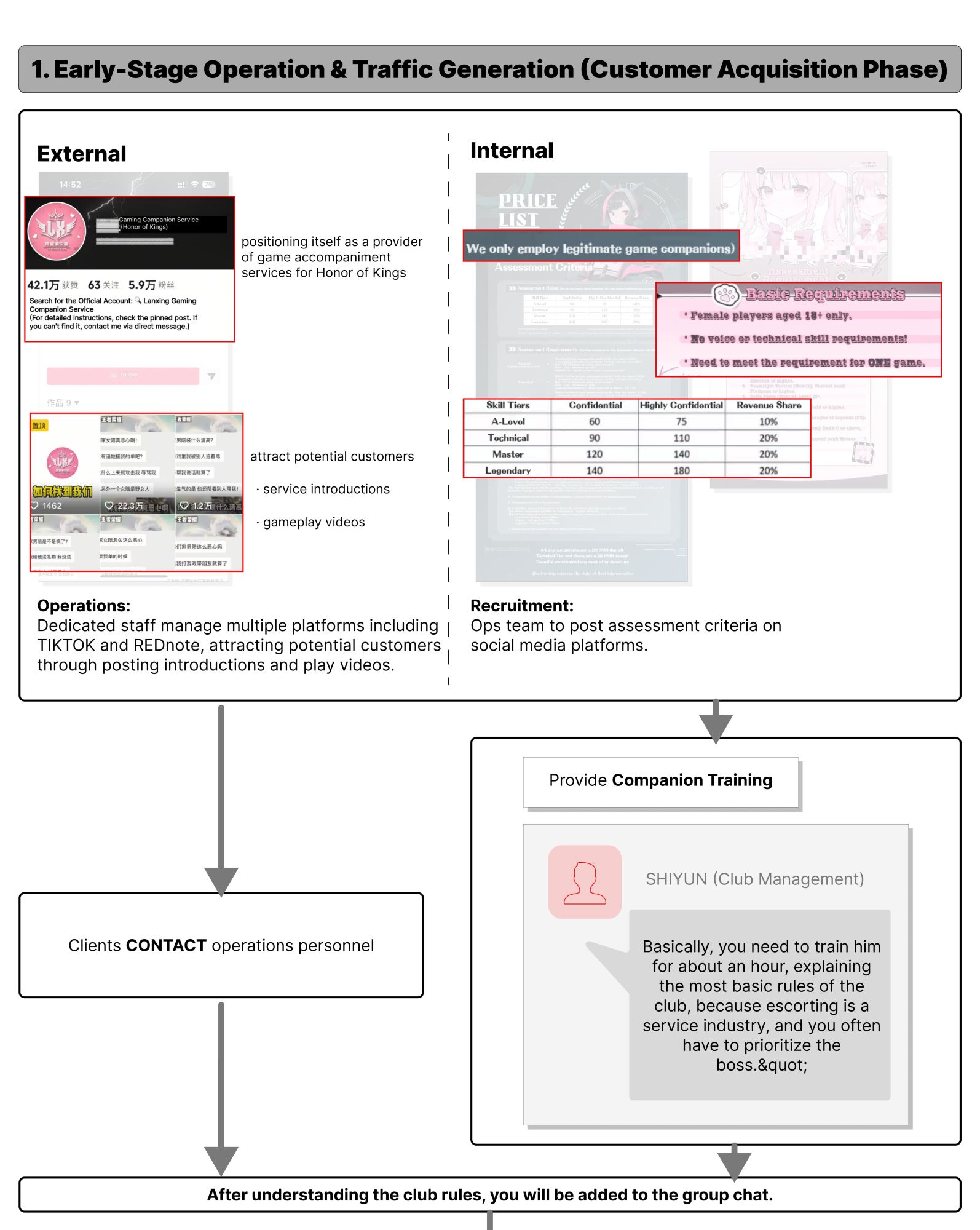} 
  \caption{Phase 1 - Early-Stage Operation \& Traffic Generation}
  \Description{A flow diagram of the customer-acquisition phase in a
  game-companionship club. Externally, dedicated operations staff use
  platforms such as TikTok and REDnote to position the club as a provider
  of game-companionship services and attract potential clients through
  service introductions and gameplay videos. Internally, the club recruits
  companions through social-media advertisements that specify eligibility
  requirements, skill tiers, prices, and revenue shares. Prospective clients
  contact operations personnel, while recruited companions receive training
  in club rules before being added to the group chat.}
  \label{fig1}
\end{figure}

Client acquisition for club-affiliated companions relies on organized operations and large-scale multi-channel promotion(\autoref{fig1}). Companion clubs set up dedicated marketing teams to publish promotional content on in-game public channels, mainstream social media (Xiaohongshu, Douyin), and gaming communities, highlighting selling points such as “professional team”, “skill guarantee”, and “hassle-free after-sales service”.

\subsubsection{Pre-operation}
\begin{figure}[htbp!]
  \centering
  \includegraphics[width=0.7\linewidth]{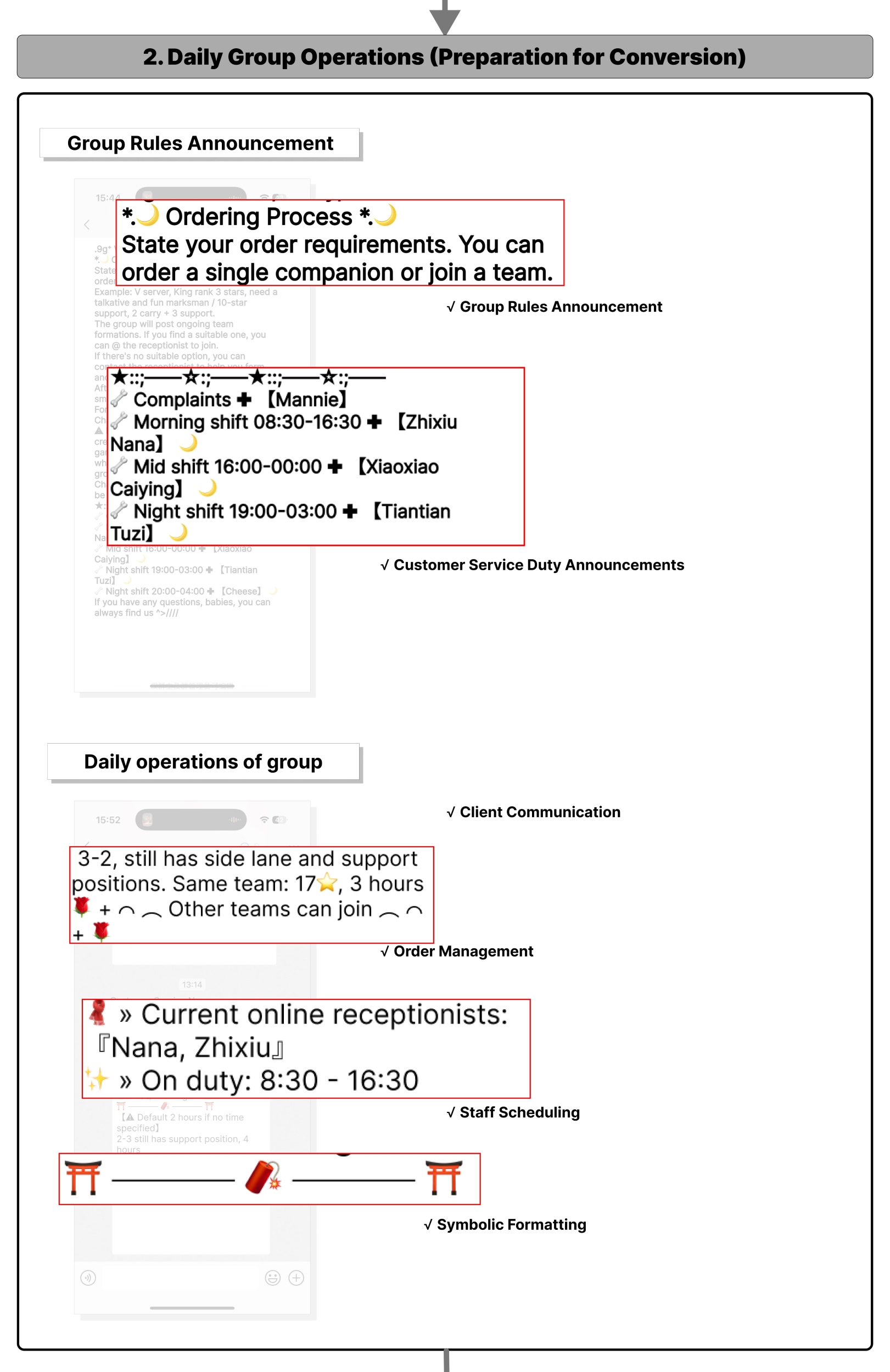} 
  \caption{Phase 2 - Daily Group Operations}
  \Description{A diagram of daily group operations that prepare potential
  clients for conversion into paid orders. Group announcements explain how
  to state order requirements, join a team or request an individual
  companion, file complaints, and identify the receptionists assigned to
  morning, middle, and night shifts. Routine group messages advertise open
  team positions, manage orders, announce which receptionists are online
  and their working hours, and use decorative symbols and emojis to create
  a recognizable communication style. The examples illustrate how the club
  combines client communication, order management, staff scheduling, and
  symbolic formatting within group chats.}
  \label{fig2}
\end{figure}

During pre-service operations, some clubs produce standardized profile cards for companions, including game ID, preferred roles/playstyles, service type, pricing, match records, and client reviews to build trust(\autoref{fig2}). As one club manager noted:“Each companion has their own profile card. Their preferred roles and main heroes are listed, along with any achievements they have earned.” (P8)The club also implements a screening process that requires companions to provide rank verification and voice samples to ensure service quality.“We require voice recordings. The club rates companions based on their voice and skills, and prices vary according to the rating.” (P5)In addition, pre-service training is provided. “The training usually lasts about one hour, covering basic club rules. Since game companionship is a service industry, we emphasize that the client always comes first.” (P8)

\subsubsection{Service Execution} 
\begin{figure}[htbp!]
  \centering
  \includegraphics[width=0.7\linewidth]{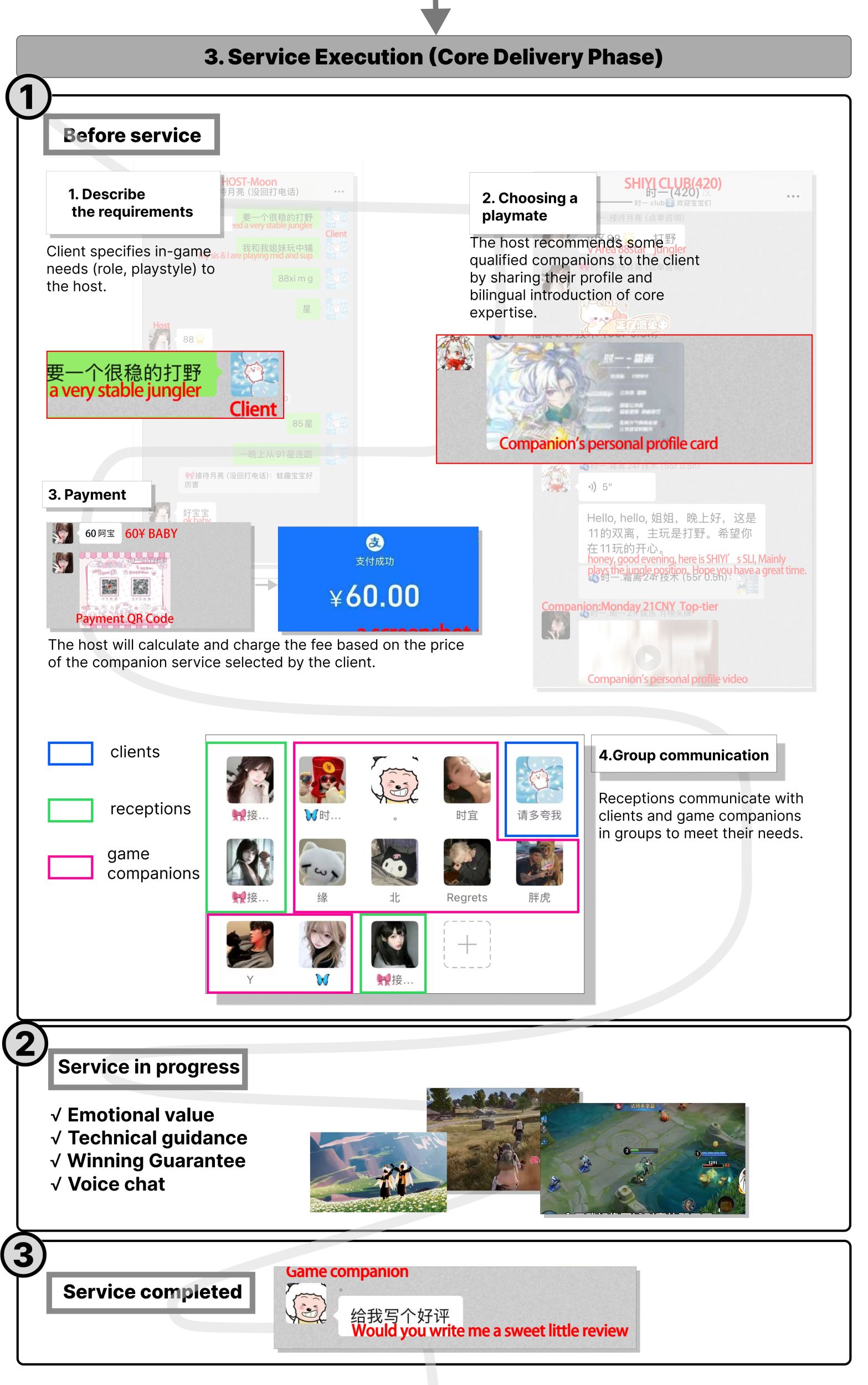} 
  \caption{Phase 3 - Service Execution (Core Delivery Phase)}
  \Description{A three-stage flow diagram of the core service-delivery
  phase. Before service, the client communicates desired in-game roles and
  play styles; a receptionist recommends qualified companions using profile
  cards and introductory materials; the client pays the calculated fee; and
  the receptionist coordinates clients and companions in a group chat.
  Color-coded outlines distinguish clients, receptionists, and game
  companions. During service, companions provide emotional value, technical
  guidance, voice communication, and assistance intended to improve the
  client's chance of winning. After service, the companion asks the client
  to leave a positive review.}
  \label{fig3}
\end{figure}

During service delivery, clubs ensure quality through standardized procedures and real-time monitoring(\autoref{fig3}).After a client places an order, customer service matches a companion based on their needs, clarifies the service duration and content (e.g., “3 ranked matches for rank-up in Honor of Kings”), and creates a dedicated chat group for tripartite coordination to avoid order skipping.“I was assigned to an order, and the client manager usually creates a small group with me, the client, and the manager. We exchange greetings first, confirm the general rank, I borrow an account if needed, and then the client sends the room ID…” (P18)Meanwhile, clubs strictly prohibit private contact between companions and clients to prevent off-platform transactions.Companions face severe penalties if caught exchanging contact information or conducting unregulated transactions outside the club.“We have clear rules: companions are not allowed to contact clients privately. If a client complains, we forfeit their withheld monthly wages and dismiss them immediately.” (P19)
This strict mechanism protects the club’s core interests while also safeguarding clients.“Some companions add clients privately, defraud them of money, and then cut off contact, or even flirt with multiple clients at once. This often happens with popular companions… A client was once defrauded of over 4,000 RMB by such a companion, excluding gifts paid for on their behalf.” (P19)

\subsubsection{Post-Service Maintenance and Client Retention}
\begin{figure}[htbp!]
  \centering
  \includegraphics[width=0.7\linewidth]{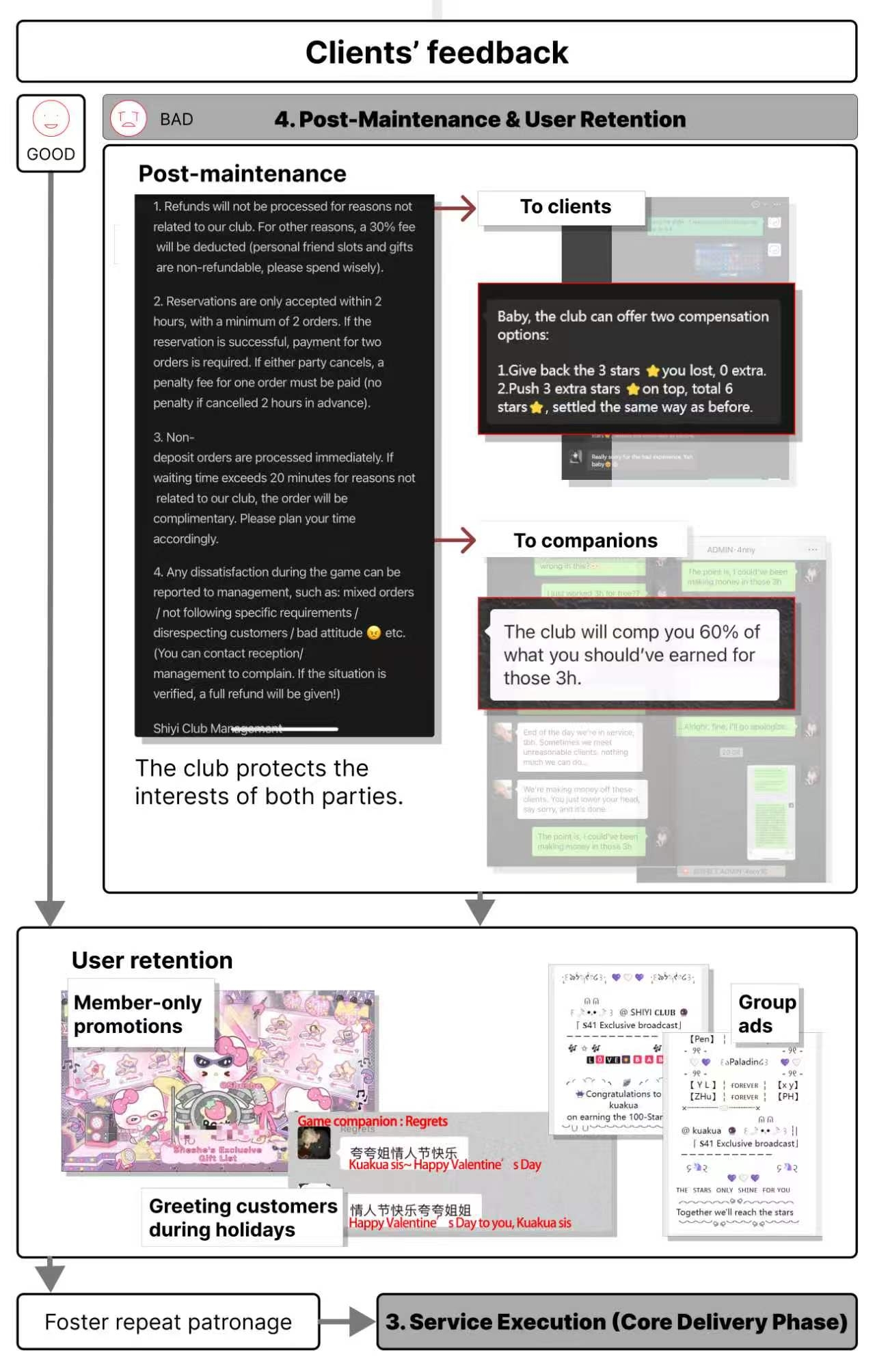} 
  \caption{Phase 4 - Post-Maintenance \& User Retention}
  \Description{A flow diagram of post-service maintenance and user-retention
  practices following positive or negative client feedback. For problems or
  dissatisfaction, the club applies refund, cancellation, complaint, and
  reservation rules and offers compensation to both clients and companions,
  thereby protecting the interests of both parties. For continued retention,
  the club uses member-only promotions, holiday greetings, and advertisements
  in group chats. These practices are intended to encourage repeat patronage
  and return clients to the core service-delivery phase.}
  \label{fig4}
\end{figure}

Clubs retain clients through systematic operations, including membership systems and exclusive events (\autoref{fig4}). More importantly, clubs assign dedicated public relations or management staff to handle in-order and post-order disputes, and adopt a two-way appeasement strategy during conflicts to balance client retention and labor stability. Clients can provide immediate feedback on dissatisfaction, and clubs usually offer compensation to maintain clients, such as fines and apology letters.
“To be honest, companions sometimes have very few rights in the club… We have to apologize to the client. Some clients won’t accept a short apology and demand a full-page one, and we’ll get fined anyway.” (P18)For highly dissatisfied clients, club PR staff will create relevant public posts  and blacklist the targeted companions within their network. On the other hand, to prevent companion turnover caused by strict management, clubs sometimes provide financial compensation and emotional comfort to companions.“We first calm the client down, then communicate and comfort the companion. If the client refuses to pay, the companion still devoted their time. So our club offers financial compensation… For example, if the win rate was unsatisfactory after one hour and the client was unhappy, we won’t force the client to pay and will cover the companion’s fee ourselves.” (P8)

\subsubsection{Core Workflow of Freelancer}
The workflow of freelancers centers on independent personal operation, with flexibility at every stage. Client acquisition relies on personal channels, such as posting service information on social media and referrals from regular clients. “I mainly post Honor of Kings companion videos on Xiaohongshu, including highlight clips and funny moments with clients. I leave my WeChat contact; after clients add me, I share my stats and prices, and they place orders if it works for them.” (P1)
Service delivery follows no fixed process and is fully negotiated between the companion and the client.
“One hour before the order, I confirm the client’s needs: whether they want fast rank-up or casual play with chatting. Then I prepare the corresponding heroes or topics.” (P6)Some freelancers build a simple personal brand on social media, or attract clients by posting gameplay videos, service details, and client feedback.
Post-service maintenance depends on personal relationship management with no systematic mechanisms.
“It mostly comes down to good chemistry. Clients order again if they’re satisfied. I send personalized greetings to regular clients on holidays to maintain the relationship, and some even keep in touch offline.” (P7)Peers also refer clients to each other.“If a client needs help, I can recommend other companions or directly connect them.” (P11)

\begin{table*}[htbp]
\centering
\small 
\caption{Differences in Work Processes between Club Game Companions and Freelance Game Companions}
\label{tab:game_companion_work_diff}
\begin{tabularx}{\linewidth}{@{}L L L@{}} 
\toprule
\textbf{Work Process} & \textbf{Club Game Companion} & \textbf{Freelance Game Companion} \\
\midrule
Customer Acquisition & Professional marketing team, cooperation with streamers, community promotion, large-scale traffic acquisition & Personal social media self-media, recommendation from regular customers, mutual recommendation among peers, scattered traffic acquisition \\
\midrule
Service Matching & Standardized matching by customer service, high efficiency & Direct personal communication, flexible negotiation \\
\midrule
Service Monitoring & Training to optimize service quality, quality is constrained & No external monitoring, entirely self-constrained \\
\midrule
Revenue Distribution & Club takes commission, game companions get 60\%-80\% of the income, stable income & No commission deduction, all income belongs to individuals, unstable orders \\
\midrule
Customer Retention & Membership system, exclusive activities, systematic maintenance & Personal relationship maintenance (e.g., holiday blessings), no fixed mechanism \\
\midrule
Risk Bearing & The club bears the risks of complaints and refunds, with more comprehensive protection of the rights and interests of both parties & Individuals bear the risk of "order default", and it is difficult to safeguard rights \\
\bottomrule
\end{tabularx}
\end{table*}

\subsection{Game Mechanics and Emotional Labor}\label{sec4.3}
The underlying differences in game mechanics serve as the core driving force behind the differentiated labor practices of game companions.The different logics of the game directly define the core competencies of the companions and the focus of the service.Voice collaboration and social compensation form the common foundation of the companion industry across all game types, with cooperative companionship and real-time emotional responsiveness as core commonalities across all service scenarios.

\subsubsection{Competitive Games}
Competitive games center on rank promotion and match outcomes. Their core mechanics—high-intensity real-time confrontation and heavy teamwork—require companions to act as both a technical backbone and an emotional buffer.
A guaranteed win rate is the fundamental requirement of such services. As respondent P18 noted: “Most clients think that if they hire a companion, I must help them win.” On the basis of ensuring basic technical performance, companions conduct real-time tactical communication via the in-game voice system, and provide timely emotional feedback and dedicated attention through frequent in-game interaction. As P19 stated:“I have to make the client feel that I am only here for them and focused on them.”Emotional fluctuations during matches demand immediate responses. When clients experience negative gameplay, companions must quickly offer comfort and redress. P19 described this practice clearly:“If the client is repeatedly killed, I say, ‘Don’t worry, I’ll get them for you.’ After defeating the opponent, I tell the client, ‘I avenged you.’” In-game voice is the essential medium for this service. P17 explicitly stated that games suitable for companionship “must have a voice chat function”, and P14 also noted that communication with clients “mostly uses in-game voice, since it’s built into the game”.Voice chat enables far more efficient tactical coordination and emotional transmission than text, making it a prerequisite for competitive companion services.

\subsubsection{Hybrid games}
Hybrid games, represented by MMORPGs, feature challenging dungeons, repetitive daily tasks, and vast open worlds. These core mechanics have created a dual demand for companions: lowering skill barriers and providing dedicated companionship.
On one hand, the high complexity and grind intensity of such games drive strong client demand for efficiency and smooth progression. As P13 noted:
“The dungeons and world bosses in Where the Winds Meet are quite demanding. Clients hire me to clear content easily and get higher ratings.”
Companions use their mechanical expertise to “save a great deal of time” for clients, helping them achieve core goals such as dungeon clears and gear progression.
On the other hand, long-form storytelling and broad social systems in open worlds often lead to loneliness in solo play. In-game guilds cannot provide dedicated companionship. As P11 stated:
“Guilds have many people, but no one focuses on you. A companion follows you the whole time and does whatever you want—it’s better than being ignored in a guild.”
Companions fill clients’ emotional void through full-time story accompaniment, casual interaction, and in-game photography. P13 added:
“Some storylines are long. Clients play with me so we can chat, and I can take photos for them. It makes them less lonely.”

\subsubsection{Entertainment-Oriented Games}
Casual social games represented by Sky: Children of the Light feature non-linear, low-guidance design and strong interpersonal interaction as core mechanics.These shift the companion’s labor focus from technical support to exploration guidance and emotional compensation, with emotional labor becoming absolutely dominant.
First, the low-guidance design and vast map create a high exploration barrier. The difficulty of finding hidden collectibles and niche scenic spots has spawned demand for guidance.As P20 noted:“Collecting Winged Lights in Sky is very time-consuming and unfriendly to impatient players, since they are numerous and scattered. That’s how this market demand gradually emerged.”Companions act as “living maps” to lower exploration costs and help clients complete check-ins and collection tasks.More importantly, interpersonal interaction is the core playability of such games. Rich two-player interactive actions create a space for simulating intimate relationships.Companions shift their service focus to high-density emotional companionship, fulfilling clients’ need for emotional compensation by creating a romantic atmosphere.P7 clearly stated:“I try to create a romantic feeling for the client in a relaxed and humorous way, letting them feel like they’re in a relationship.”Companionship is the very essence of this game genre. As P7 summarized:“After playing Sky for a long time, you’ll realize its playability exists entirely because of companionship. The quests and tasks are all fixed.”

\begin{figure}[htbp!]
  \centering
  \includegraphics[width=1\linewidth]{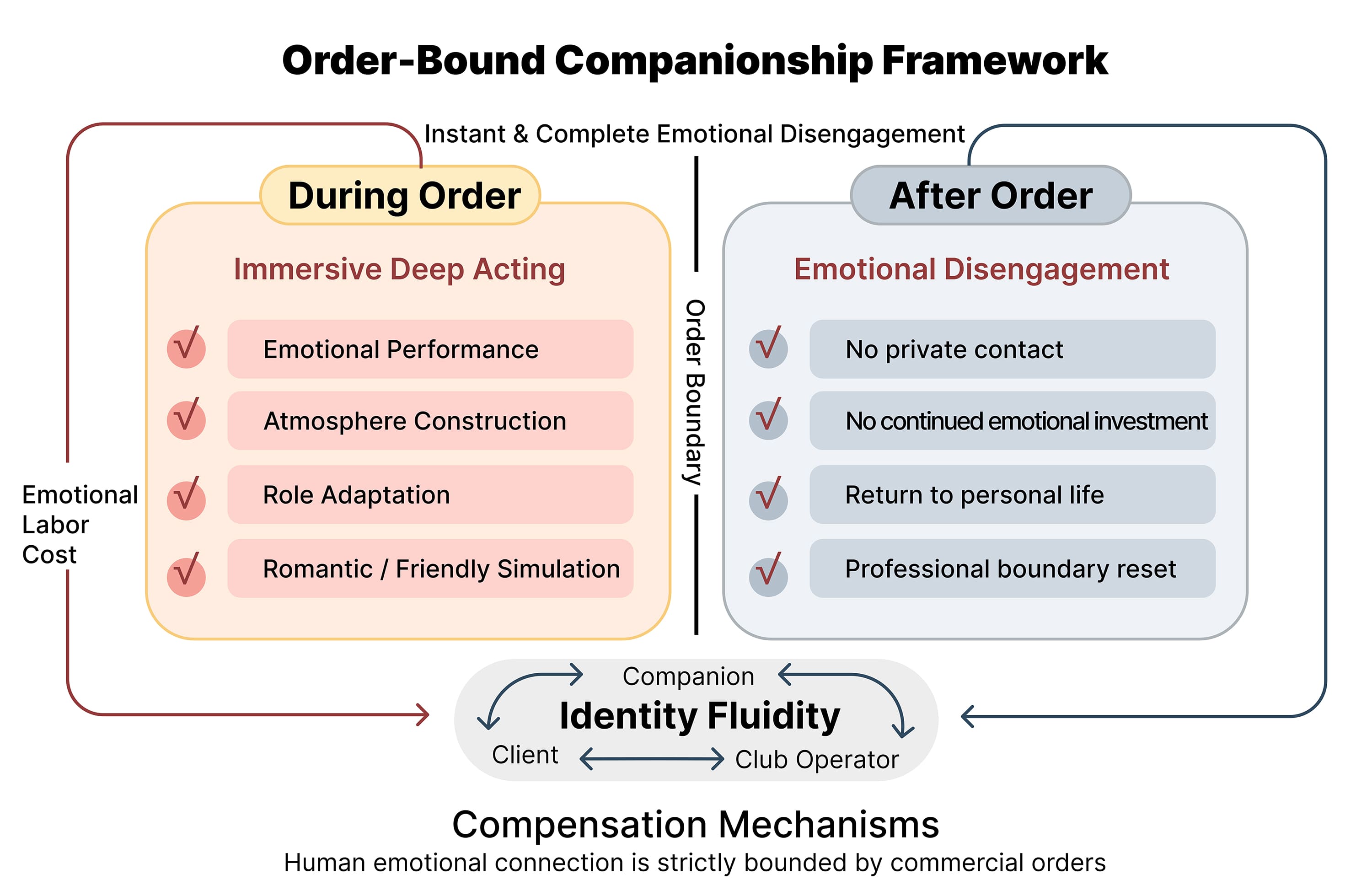} 
  \caption{The "Order-Bound Companionship" Framework}
  \Description{The Order-Bound Companionship Framework contrasts companions'
  emotional practices during and after a paid order. During an order,
  companions perform immersive deep acting through emotional performance,
  atmosphere construction, role adaptation, and simulated romantic or
  friendly interaction, which creates emotional-labor costs. At the order
  boundary, they shift immediately toward emotional disengagement by avoiding
  private contact, ending further emotional investment, returning to personal
  life, and resetting professional boundaries. Arrows also show identity
  fluidity among the roles of companion, client, and club operator. Together,
  these mechanisms strictly confine human emotional connection to the
  duration of commercial orders.}
  \label{fig5}
\end{figure}

\subsection{The Social Network Landscape Centered on Game Companions}\label{sec4.4}
As providers of emotional value, game companions act as the core connectors of the industry’s social network. This network is interwoven by three pivotal relationships: game companion–client, game companion–game companion, and game companion–club (\autoref{fig5}). Driven by commercial profitability, the industry features a coexistence of genuine emotion and transactional interest.

\subsubsection{Game Companion–Client: The Core Bond of the Social Network}
Fundamentally a commercial service, this relationship defines the scale of emotional provision through scenario-based boundaries. At its core, companions deliver "companionship-style services" and emotional value, ranging from technical gaming support to emotional labor like bedtime soothing, chat accompaniment, and simulated romance. Although both sides often aim to move beyond hierarchical service ties to pursue equal, friend-like interactions, one client stated:
"I don’t want game companions to pander to me. It makes the relationship feel deeply unequal. I just want someone to play with me... I hope we can get along as friends, without anyone feeling tense." (p. 7)

Yet, the inherent service dynamic keeps clients in a dominant position. The selection process is asymmetrical: initially, companions have little information about their clients but must maintain consistent emotional performance to meet expectations. Highlighting this vulnerability, one companion recounted her experience with a difficult client:
"I only found out what he was like after playing with him, but I couldn’t bail on the order halfway—because bailing would get me a complaint... Next time, if I get assigned to this same client again, I’ll either turn down the order or check with the manager first." (p. 19)

\subsubsection{Scenario-Based Emotional Boundaries} 
To balance commercial emotional performance with practitioners’ boundary protection, the industry has formed a clear consensus around "order-bound companionship" (\autoref{fig5}). This is particularly prominent among entertainment-based companions providing romantic services, though it applies across all categories. For example, among companions on \textit{Sky: Children of the Light} who specialize in romance-themed services, a shared consensus prevails: “During the order, I’m your lover; outside the order, I’m not.” (p. 7)

This statement clearly demarcates the emotional boundary: the end of the order serves as the dividing line. During an order, to boost client spending, companions rationally accept ambiguous advances and form virtual emotional bonds with romantic undertones. Once the order ends, these virtual connections are strictly severed. One companion noted:
“Every companion who takes romance-themed orders is mentally prepared for clients to develop feelings... If they like you, they’ll spend more on you. I just acknowledge their feelings without liking them back.” (p. 20)

This boundary is heavily enforced for club-affiliated companions. Club managers explicitly ban private contact between companions and clients (p. 5, 7, 9, 18, 20, 21). Private communication or romantic relationships result in severe penalties, including dismissal for companions and deducted balances for clients. This protects both parties and prevents private emotional or financial disputes. A club receptionist explained:
“Once a companion dates a client... it hurts their business—they can’t reasonably take orders from other clients of the opposite gender... For clients, there’s a risk: some popular companions deliberately solicit money, break up after getting enough, or lead multiple clients on at the same time.” (p. 19)

In rare cases, however, genuine friendships develop beyond the service relationship:
“I once had a wealthy client from the US; we were purely friends... With no ambiguity, getting along as friends felt really comfortable.” (p. 17)

\subsubsection{Game Companion–Game Companion: The Dynamic Peer Network}
As the most dynamic dimension of this social network, peer relationships among game companions are characterized by a blend of competition and symbiosis over client resources. 

On the competitive front, companions vie for orders both explicitly and implicitly. This sometimes manifests in "order-snatching," where a companion unilaterally takes an order without the client's prior consent to switch.
“The client specifically @ed me and asked me to join... I just didn’t check my phone for that one minute. A minute later... I saw another game companion directly join the room. The client didn’t say to replace me, but that person just went in on their own.” (p. 16)

Symbiotically, companions form mutual support systems for resource exchange, emotional relief, and skill development. High-volume companions frequently refer clients to peers:
“I also recommend other companions to my clients... I know they won’t only order me. And since they’re my friends, I don’t mind referring them.” (p. 9)
“For example, I get a lot of orders, so other companions think I have more client resources and can influence who the client chooses... So sometimes, they also refer their own clients to me.” (p. 11)

When facing unreasonable client demands, companions relieve stress by privately venting to peers. As one companion mentioned regarding management siding with a client:
“The higher‑ups just think the client is always right... Then I’d talk to other companions privately, say how unreasonable the client was, and they’d agree with me. That way, I don’t feel stressed anymore.” (p. 18)

Additionally, they consult more skilled peers to refine their gaming and emotional service abilities, though these exchanges typically only happen between close friends:
“Usually, we share when chatting with friends, or if we want to improve our skills, we ask people we think are better than us.” (p. 20)

\subsubsection{Game Companion–Game Companion Club: The Organizational Core}
A governance relationship marked by interdependence and restraint binds companions and clubs. Serving as employment and support platforms, clubs form a reciprocal dynamic with workers: companions rely on the club for client resources and protection, while the club regulates service conduct and resource access.

\textbf{Dependence (Risk Avoidance and Resource Security):} 
Dependence manifests in resource security and risk avoidance. Clubs control core client and profile resources. Although most companions are not permanently bound to a single club, their client streams are deeply tied to it. A former club manager noted: 
“When you sell a club, clients are sold as assets. Clients, companion profiles, and membership size all form the valuation of the club.” (p. 7) 
Regarding private order-taking, she added: “Most of the time, clubs turn a blind eye to companions switching clubs, but the bottom line is: you cannot take clients away with you.” (p. 7)

When facing harassment or unreasonable demands, companions can report to the club. The club intervenes to reject such requests, issue public announcements, and blacklist problematic clients (p. 21). Senior administrators mediate unreasonable behaviors, shielding companions from direct emotional conflict (p. 4).

\textbf{Restraint (Rule Enforcement and Conduct Management):} 
Restraint is reflected in rule enforcement and conduct management. Clubs impose clear rules on emotional delivery. For instance, when serving a client together, companions must “focus on the client, limit cross-companion conversation to no more than five sentences, and keep topics centered on the client” (p. 19).

Companions who violate rules or receive repeated complaints face penalties including wage deductions, deposit forfeiture, suspension, and dismissal (p. 2, 4, 5, 9).

Clubs maintain hierarchical systems with detailed promotion rules. As noted, promotion via "arena challenges" is common:
“Lower-ranked companions can challenge higher-ranked ones... The two treat each other as clients and compete on professional skills… Some act aggressively, testing the opponent’s stress tolerance and responsiveness. One memorable line was: ‘You say you love me—what gives you the right to ask me to love you back?’” (p. 20)

Such simulations of client interaction effectively screen companions and evaluate their emotional service capabilities. Finally, highly popular companions form a delicate, two-way restraint with the club: while they attract significant traffic, an aloof service attitude or poor attendance can harm club operations. Consequently, the club still enforces attendance rules and service standards upon them (p. 19).
\subsection{Role Positioning and Identity Mobility of the Three Core Actors from the Perspective of Emotional Labor}\label{sec4.5}
Emotional labor constitutes the core of game companions' work. Within this service ecosystem, three central actors—game companions, clients, and game companion clubs—form interconnected, mutually supportive roles with distinct focuses. Clubs, in particular, act as a crucial emotional buffer between companions and clients. Furthermore, these roles are not static. Significant identity mobility allows individuals to shift between groups, enriching the industry's emotional labor ecology.

\subsubsection{Game Companion}
As the core providers of emotional labor, game companions bear the full-process responsibilities and psychological costs of emotional expression, suppression, and exhaustion.

\textbf{Emotional Expression:} Companions must accurately perceive clients’ emotional needs. For instance, an \textit{Honor of Kings} companion (p. 16) mentioned she would help clients recover lost ranks or switch teammates to ease their frustration. She also acted cautiously when sensing clients were troubled by real-life issues: 
“If I’ve played with him many times, I can tell his mood is different... so if he seems down, I might not dare talk much and will be very careful.” 
Companions also adapt their service styles to the game genre. In competitive games, they foster a sense of achievement by helping clients win, whereas in social exploration games, they focus on relaxation and relieving emotional distress (p. 7).

\textbf{Emotional Suppression:} During service, companions must strictly suppress their own negative emotions. They cannot show dissatisfaction toward unreasonable demands and are forbidden from bringing personal peer conflicts to clients (p. 19). They must also perform emotional masking when a client is in a bad mood (p. 14). To separate personal feelings from work, some companions treat their service as “role-playing.” When asked about the difference between romance-themed and ordinary orders in \textit{Sky: Children of the Light}, one companion stated: 
“I feel nothing. To me, it’s just like role-playing; there’s no real emotional reaction.” (p. 20)

\textbf{Emotional Exhaustion:} Providing sustained emotional value and acting as an “emotional dumping ground” places a heavy psychological burden on companions. One companion explained: 
“Sometimes it really feels like a burden. Some clients don’t want to see me interact with other clients... They may expect a one-on-one relationship, but I serve many clients, who all share things with me... It can get tiring.” (p. 16) 
Popular full-time companions may even succumb to a numb state of “emotional emptiness”: 
“After taking too many orders, you’re constantly giving emotional value to others until you feel numb... and become emotionally empty.” (p. 7)

\subsubsection{Client} 
As the primary recipients of emotional value, clients' needs dictate how companions perform their emotional labor. However, this relationship is not a one-way street; clients also offer reverse emotional value, which serves as a key job motivation for companions. Almost all female companions described themselves as homebodies with ample spare time. The clients’ need for company aligns perfectly with the companions’ own desire to “have someone to play with,” providing them with reciprocal emotional satisfaction (p. 3, 7, 11, 13, 16, 17, 19, 20). This two-way emotional interaction introduces flexibility into their labor dynamic.

\subsubsection{Game Companion Club} 
Clubs function as the crucial emotional labor buffer zone between companions and clients.

\textbf{Risk Buffering:} The club mediates extra-order conflicts. If companions face entanglement, unreasonable behavior, or harassment, senior managers or receptionists intervene. This prevents direct emotional confrontations and shields companions from non-service disputes. Recalling verbal harassment from a male client, one female companion noted: 
“After we lost a game or two... He replied, ‘It’s fine—just call me baby next time.’ I was stunned... Then the receptionist stepped in, saying this counted as a sweet-talk order and required extra payment. Once money was mentioned, the male client usually went quiet.” (p. 18)

\textbf{Governance and Management:} The club regulates emotional labor through clear rules and penalties, such as strictly banning private contact and romantic relationships with clients (p. 5, 7, 9, 18, 20, 21). It also enhances companions’ core capabilities via hierarchical "arena challenges" (p. 19) and conversational skills training (p. 7). Furthermore, restrictions placed on highly popular companions ensure consistent service quality across the club, thereby sustaining the industry's basic operational order.

\subsubsection{Identity Mobility Among the Three Core Actors}
Notably, the identities of these three core actors are not fixed; they exhibit significant mutual mobility. Individuals shift roles based on personal needs and career development—most notably, transitioning from companion to client.

After providing long-term emotional support to others, full-time companions often develop a psychological state of “habitually needing company.”
“As a relatively popular game companion... I’m usually busy with orders. So when I suddenly have free time, I feel very uncomfortable. For a period, I had to order other companions myself to relieve the stress of not talking to anyone, because I’d grown used to having someone on a call...” (p. 7)

Following extensive emotional output, these workers struggle with solitude and actively become clients to purchase emotional value for themselves. This reversal—from emotional labor supplier to demander—provides them with essential emotional compensation and relief.

\section{Discussion}\label{sec:Discussion}

\subsection{Order-Bound Companionship: From Individual Strategy to Institutional Mechanism}
Classical emotional labor scholarship argues that commodifying affective connection undermines workers' ability to distinguish genuine feeling from performance, producing what Hochschild\cite{hochschild1979emotion} calls emotional estrangement. The extension of this framework to digital contexts reinforces the concern: studies of cam workers and digital sex workers document how "always-online" platform pressures continuously erode personal boundary strategies, forcing workers to choose between protecting themselves and remaining economically competitive \cite{schneider_managing_2026, hamilton_risk_2022}. The concept of bounded authenticity \cite{bernstein_meaning_2001} offers a partial counterpoint—commercial transactions themselves function as implicit delimiters of intimacy—but this delimiting function in adjacent industries remains informal and individually negotiated, making it structurally fragile \cite{sanders2002condom}.

Our findings complicate this picture \autoref{fig5}. \autoref{sec4.4} shows that the Chinese game companionship industry has developed an implicit industry rule around what we term "Order-Bound Companionship"; this is not a personal coping strategy. Within orders, companions engage in full "deep acting"—accepting ambiguous advances, sustaining romantic atmospheres, and attending closely to clients' emotional states. Once the order ends, emotional connection is structurally severed: all sampled clubs explicitly ban private contact, enforcing this through financial penalties and dismissal \autoref{sec4.4}.

This pattern aligns with Bernstein's \cite{bernstein_meaning_2001} "bounded authenticity"—the commercial transaction delimits the emotional encounter—but extends it in two ways. First, where Bernstein describes an emergent, transactional convention, "Order-Bound Companionship" is institutionalized: the boundary is encoded in platform architecture, club rules, and the natural endpoint of a gaming session. Second, where prior digital intimate service research shows boundary strategies being systematically eroded by platform competition \cite{schneider_managing_2026}, the "Order-Bound" mechanism removes the individual from having to negotiate and defend their own boundary at all. The club and order structure do it on their behalf. This resonates with Sather's \cite{sather_persona-mediated_2026} concept of "persona-mediated dissociation"—adaptive persona use requires structural support for role exit—and suggests that when such support is institutionally provided rather than personally maintained, the psychological cost of the labor form is substantially reduced.

However, the "Order-Bound" mechanism is not simply a worker-protective device. As \autoref{sec4.4} also shows (P19), club prohibitions on private contact serve organizational interests: they protect the companion's market value, prevent off-platform transactions, and maintain service quality predictability. This dual function—protecting workers while serving managerial imperatives—reflects a broader pattern in platform-mediated emotional labor and warrants a dialectical rather than celebratory reading. Nevertheless, compared to the individually-maintained and relatively unstable boundary strategies in adjacent industries, the institutionalized mechanism observed here reflects a noteworthy logical shift: here, it is more the service structure itself that provides boundary support, rather than relying entirely on the worker's continuous input of personal energy.

\subsection{The Tripartite Governance of Relational Digital Labor}
Gig economy research has predominantly theorized labor governance as bilateral \cite{gandini2019labour}: platforms exert algorithmic control over workers through performance metrics, ratings, and income structures. In the Chinese context, prior scholarship characterizes platform intermediaries (guilds and MCN organizations) primarily as manipulation structures \cite{liu_zhibo_2023}, discursive authority mechanisms\cite{xiao2025institutionalizing}, or ambivalent support-and-control institutions whose governance workers experience as coercive \cite{tsang_hope_2025}. Zhao \cite{zhao_fragmented_2023} identifies these intermediaries as nodes of fragmented control at the labor-process level, governing order allocation and revenue distribution.

Our findings (\autoref{sec4.4} and \autoref{sec4.5}) add a dimension to this picture that prior accounts do not capture: the emotional governance buffer. Clubs in this ecosystem perform three functions that operate at the level of individual service interactions rather than structural labor control. First, real-time risk absorption: when companions face harassment or unreasonable client demands, club managers and receptionists intervene before the companion must engage directly (\autoref{sec4.5}). This directly addresses the vulnerability identified in Shen’s prior study of Bixin companions\cite{shen_labor_2021}, which found that female companions had few countermeasures against harassment. Our data show that club affiliation provides what freelance companions lack. Second, bilateral emotional regulation: rather than simply adjudicating complaints, clubs manage the emotional states of both parties, drawing on relational knowledge of individual clients and companions that algorithmic systems cannot replicate \autoref{sec4.2}. Third, capacity cultivation: through arena-style training, hierarchical mentoring, and rule-enforced skill development \autoref{sec4.4}, clubs systematically build companions' affective service capacity beyond what individual experience accumulates.

This stands in contrast to existing accounts of Chinese intermediaries. Where Liu et al. \cite{liu_zhibo_2023} emphasize guilds as algorithmic performance orchestrators, and Xiao et al.\cite{xiao2025institutionalizing} focus on their discursive authority over workers, our data reveal a dimension that neither framework addresses: emotional mediation at the moment of affective encounter between worker and client. Where Tsang and Wilkinson \cite{tsang_hope_2025} document the structural ambivalence of guild relationships (support and control co-existing), our data show this ambivalence playing out in real-time service interactions rather than only in contractual arrangements. The two perspectives are complementary rather than competing: Zhao \cite{zhao_fragmented_2023} and others describe the macro-structural control logic; our findings add the micro-level emotional governance function.

The identity mobility documented in \autoref{sec4.5} enriches this picture further. Companions who develop a habitual need for company become clients; experienced companions become club managers. P7's account—having to hire companions herself because she could no longer tolerate silence—also illustrates how intensive emotional labor output reshapes practitioners' affective needs, thereby sustaining demand within the same ecosystem.

In summary, we position this study as a context-specific case study: the "Order-Bound" mechanism and the tripartite emotional governance structure are jointly shaped by China's regulatory vacuum, the multi-platform ecology (Bixin, WeChat, Douyin, Xiaohongshu), and cultural expectations around emotional service. Whether this configuration would emerge in other regulatory or cultural contexts remains a direction for future discussion.

\subsection{Design Insight}
Combining our findings with existing literature, we propose design implications for game companionship and related platforms across three core functional domains: game mechanics design, platform system architecture, and community governance and worker protection.
\subsubsection{Game Mechanics Design}Prior HCI and games research has extensively examined how game mechanics shape social narratives: from tactical communication mechanisms in highly competitive MOBAs and FPS games \cite{xu_sociable_2011,lee_less_2025,kou_playing_2014}, to casual social interaction patterns in titles like Animal Crossing \cite{tong_players_2021}. These studies emphasize how mechanics function as a social language, choreographing emotional distance between players. However, when transitioning to the context of commercialized game companionship, we argue that game mechanics transcend general social facilitation; they become the core infrastructures of emotional labor. Unlike prior design recommendations aimed at fostering general player connection, our findings suggest that in paid companionship, mechanic design must actively support the service providers' emotional efficiency and psychological boundary management. 

Voice-First Architecture as Emotional Delivery Infrastructure. While Walther\cite{walther1994interpersonal} established that computer-mediated communication (CMC) carries affective content, prior game studies have typically treated in-game voice primarily as a tool for tactical coordination or an optional layer for social presence. In contrast, our data \autoref{sec4.3} reveals that in paid companionship, built-in voice is the non-negotiable foundation for real-time emotional delivery (P14, P17). It provides an emotional immediacy that text cannot replicate, thereby reducing service friction. Therefore, rather than treating voice as an optional add-on, we recommend that designers embed it as a core affective affordance. This requires moving beyond basic audio transmission to include companion-specific voice modes, low-latency native channels, and granular privacy controls that empower companions to manage interaction boundaries within a single session.

Embodied Interaction Mechanisms as Shared Emotional Labor. Traditional emotional labor scholarship\cite{bono200512,grandey2000emotional} highlights that "deep acting" incurs high psychological costs because it demands the worker's internal emotional alignment. Building on this, our findings \autoref{sec4.3} demonstrate how game mechanics can mitigate these costs: embodied mechanisms like hugging and synchronized flying in Sky: Children of the Light actively shift the emotional burden from the companion’s individual verbal performance to co-participatory physical actions (P7). To alleviate the individual performance burden of companions, we recommend that social games intentionally incorporate two-player synchronized mechanisms. However, unlike traditional social game designs that universally encourage deep player attachment, our study notes that such mechanisms in commercial companionship can easily lead to client over-investment. Thus, our core recommendation differs: such intimate in-game mechanics must be strictly coupled with platform-level boundary enforcement (i.e., the "Order-Bound" mechanism) to ensure that companions can safely and smoothly execute role exit at the end of an order.

\subsubsection{Platform System Architecture}
Implementing pre-order service contract modules is essential for establishing a baseline of mutual consensus. \autoref{sec4.2} shows that the clubs' pre-game tripartite group chats (P18) establish this consensus for the "Order-Bound" boundary; freelance companions lacking this process face higher risks of expectation misalignment (P19). Platforms should construct mandatory pre-order contract modules within the service workflow for both parties to clarify service types, durations, and behavioral norms before playing. This formalizes informal practices into platform design features, addressing the structural vulnerability identified in prior digital intimate work where workers individually bear the costs of ambiguous boundaries \cite{hamilton_risk_2022, palatchie_currying_2025}.

Developing structured worker incentive systems beyond simple ratings is crucial for acknowledging the full scope of companion labor. Relying solely on client ratings reproduces the algorithmic disciplinary dynamics described by Srnicek\cite{srnicek2017platform}. Companions' invisible labor—such as persona construction and rule inference—remains systematically ignored and uncompensated (\autoref{sec4.5} )\cite{kojah_dialing_2025, ming_i_2023}.  Platforms should design multi-dimensional incentive structures that capture what ratings miss: acknowledging long-term client retention, awarding skill development badges for completing training milestones, and incorporating peer referral points that formalize the symbiotic knowledge sharing documented in \autoref{sec4.4} (P9, P11). This shifts platform design from pure output metrics toward acknowledging and compensating the entire labor ecosystem.

The provision of worker-facing algorithmic transparency dashboards can significantly alleviate the unpaid energy companions expend decoding platform rules\autoref{sec4.5}. As Liu et al. \cite{liu_zhibo_2023} observed, platform opacity forces workers to guess, positioning intermediaries as algorithmic experts. Platforms should provide intuitive transparency interfaces that explicitly detail recommendation mechanisms and account suspension criteria. This directly reduces the misguided community labor identified by Kojah et al. \cite{kojah_dialing_2025} and lowers independent companions' dependence on clubs as algorithmic mediators.

\subsubsection{Community Governance and Worker Protection}
Establishing tiered enforcement of community-wide client behavior standards is crucial for safeguarding all workers. Clubs currently protect companions from harassment through real-time interventions and blacklists(\autoref{sec4.5}, P18, P21), whereas freelance companions lack such institutional shelter—a common structural vulnerability in digital intimate work\cite{shen_labor_2021, hamilton_risk_2022}. Platforms should encode behavior standards into enforceable platform policies and establish a tiered response system (warnings, restricting access to romance-themed orders, and community-wide permanent blacklists). This transforms informal club-dependent protection into a platform-wide infrastructure accessible to all companions.

Integrating tiered conflict mediation systems with human escalation nodes addresses the emotional complexities of relational service disputes. Clubs' manual bilateral placation strategies (\autoref{sec4.2}, P8) currently navigate these complexities, preventing companions from being trapped in difficult dilemmas (P19). Platforms should integrate tiered conflict mediation directly into the service architecture: automated processing for simple technical complaints, and human escalation for emotionally sensitive order types, accompanied by a penalty-free 24-hour mediation window. This formalizes the protection currently obtained via club intervention and echoes calls for "human-in-the-loop" governance in complex affective service contexts\cite{liu_zhibo_2023}.

\subsection{Limitation}
\subsubsection{Representativeness of the Sample and Research Background} This study only interviewed game companions in mainland China. If interviews were conducted with Western game companions and their clients, the findings might reveal a labor process more focused on skill instruction, tactical analysis, and performance evaluation—given that foreign platforms (such as Legionfarm’s professional coaching model) place greater emphasis on technical competence and professionalism. Meanwhile, Western clients’ demand for “emotional contribution,” as well as companions’ perceptions and strategies toward it, may differ from those in China, likely prioritizing “positive feedback” over deep emotional bonding or social companionship.

The research participants mainly consist of young people aged 18–25, which stems from the unsustainable and precarious nature of game companionship itself. Most companions describe it as a “youth-based career”—as age increases, declining reflexes make it difficult to meet technical requirements. However, elderly service users(clients) undoubtedly exist. Interviews with elderly game companion clients could uncover an entirely new landscape of needs. They may not pursue competitive ranks, but instead seek pure companionship or social recognition. This will expand our understanding of the “service recipients” and “value connotations” of game companionship.

In addition, most respondents were part-time game companions, which reflects the high flexibility that draws the majority of companions to work part-time. However, we also found that compared with full-time companions, part-time companions tend to have blurred work–life boundaries, lower professional identity, and more pragmatic expectations of economic returns. Their labor motivations and practical strategies may differ significantly from those of full-time companions.

Future research should expand the sample scope and adopt a cross-platform comparative approach to enhance generalizability.

\subsubsection{Static Research Design}As a cross-sectional study, this research only captures the current state of the industry and does not longitudinally track identity transitions or market changes (e.g., game popularity, policy adjustments), thus failing to reveal the evolutionary logic of labor practices.

\subsubsection{Limitations of the Interview Method}
This study relies primarily on semi-structured interviews. Although interviews allow for an in-depth exploration of practitioners’ subjective narratives and psychological perceptions, they have limitations in capturing the embodied nature of game companion labor. The actual working conditions, immediate emotional regulation, and subtle interaction strategies of companions are often deeply embedded in real-time gaming contexts.
Purely verbal retrospection may lead participants to omit highly tense, non-verbal details of emotional labor during gameplay due to memory bias or professionalized rhetoric. Therefore, this study may not fully capture the real-time physiological and psychological responses of companions under extreme gaming conditions.
In future research, we plan to introduce methods such as participant observation. By having researchers act as “clients” and genuinely participate in matches, we can not only observe in situ how companions perform the de-laborization of labor through game mechanics, but also establish deeper trust via immersive interaction. This will enable us to obtain more authentic and contextually grounded interview data after gameplay.

\section{Conclusion}\label{sec:Conclusion}
Grounded in the specific socio-cultural context of China's platform gig economy, this study explores the relational dynamics and emotional labor practices of professional game companions. We offer a micro-level perspective on how intimate boundaries and stakeholder networks are navigated in digital relational labor. Our observations illustrate that companions rely on an institutionalized “order-bound” mechanism—engaging in immersive deep acting during paid sessions, followed by complete emotional disengagement post-order—to manage their psychological well-being. Additionally, we highlight a tripartite network involving companions, clients, and club management, defined by scenario-based performances, competitive yet symbiotic peer relations, and interdependent governance. Within this ecosystem, identities remain fluid, with individuals frequently transitioning between roles driven by shifting emotional needs and resource accumulation.

Drawing on these insights, we suggest several design directions to better support workers in relational digital labor. First, game mechanics—such as voice-first architectures and embodied interactions—should be intentionally designed not just for general socialization, but as affective infrastructures that facilitate emotional delivery while strictly supporting safe role exits. Second, platform systems could better protect workers by formalizing pre-order contract modules and implementing multi-dimensional incentive structures that recognize invisible labor beyond simple rating metrics. Finally, integrating human-in-the-loop conflict mediation and community-wide behavioral standards can help institutionalize the boundary protections currently managed informally by clubs, offering a structural pathway to mitigate the psychological burden and emotional exhaustion inherent in commercial companionship.

\bibliographystyle{ACM-Reference-Format}
\bibliography{references}

\newpage

\appendix
\section{Interview Questions}\label{sec:Appendix}
\subsection{Warm-up Questions}
\begin{enumerate}
    \item Have you played games a lot before becoming a game companion? What's the biggest difference between playing games for fun back then and playing as a game companion now?
    \item What's the difference in how you feel about your esports career, playing games on your own, and working as a game companion?
    \item What's your general opinion of the game companion industry?
    \item Is there a clear hierarchy between popular game companions and regular ones? Do some look down on others?
    \item Have you ever met clients or other game companions in person?
    \item What are the possible career paths for a game companion?
    \item Will teams promote romantic experience services as a way to grow their business?
\end{enumerate}

\subsection{Basic Information}
\begin{enumerate}
    \item How long have you worked as a game companion? What main types of games do you do, and what kind of game companion are you (skill-based/entertainment-based)?
    \item Are you a full-time or part-time game companion? How many hours do you spend on this job every day? If you're part-time, what's your main job? If you're a manager, tell us the age range of your team's companions and the ratio of full-time to part-time ones.
    \item How do you set your prices, and what's your price range? Do platforms or clubs take a cut of your earnings? What are the specific rules?
    \item How did you start working as a game companion, and what's the main reason you do this job?
\end{enumerate}

\subsection{Order Taking and Client Management}
\begin{enumerate}
    \item What are your main platforms or channels for getting orders, and what percentage of your orders come from each? How do these platforms differ in how you find new clients and the types of clients they have?
    \item Do you choose different order platforms for casual MMO games and competitive games? How do you post your services and deal with payments on non-traditional platforms?
    \item What makes you better than other game companions? How is the emotional support from an entertainment-based game companion different from that from a friend?
    \item Do you run social media accounts or have a fixed personal image for your job? Do you have regular clients, and how do you keep in touch with them?
    \item What are the recruitment and assessment standards for game companions, and what do you learn in pre-job training? How does the club find new clients and help companions build their personal image? Why was the club founded in the first place?
\end{enumerate}

\subsection{Service Content and Delivery}
\begin{enumerate}
    \item What specific things do you do in your game companion service? Is there a set process for your service?
    \item How do you tell how a client is feeling, cheer them up, and deal with their bad moods? Is this emotionally tiring for you?
    \item How do you feel after finishing each service? Does it have to do with the type of game or the platform you got the order from?
    \item How do different platforms differ in their service rules and guarantees (like money protection and dispute solving)? How do these differences affect your work experience?
    \item How do you balance the relationships between game companions, clients and the club? What's the difference in service requirements between top clubs and ordinary ones?
\end{enumerate}

\subsection{Emotional Labor and Relationship Boundaries}
\begin{enumerate}
    \item Do you have to act in a certain way or show specific emotions when providing services? How do you deal with clients' bad moods and inappropriate behaviors?
    \item Besides spending your emotions on the job, what else do you need to put in for this work?
    \item What do you do when you can't meet a client's demands, and how do you feel about it?
    \item Is your relationship with clients more like a service relationship, a friendship, or something else? Does this relationship change the more you work with a client?
    \item Do you ever feel possessive of your clients? How do you feel when a client likes you the most, or when they switch to another game companion? Have you ever had real feelings for a client, and if so, what did you do?
    \item Does the team have any rules about in-game romantic relationships or adding clients as game friends? How do you deal with it when a companion or a client gets upset? What punishments are there for companions who break the team's rules?
\end{enumerate}

\subsection{Industry Interaction and Competition}
\begin{enumerate}
    \item Do you compete with other game companions, or do you help each other out? How do you interact with them when playing games and in daily life?
    \item Is the competition with other game companions fiercer on some platforms than others? Does this change how you get along with other companions?
    \item When two companions serve one client at the same time, have you ever seen the two companions chat with each other and ignore the client? How can this be avoided?
    \item Do you work well with the platform or club? Do their rules and support meet your needs for the job?
\end{enumerate}

\subsection{Career Perception and Experience}
\begin{enumerate}
    \item What are the main pros and cons of being a game companion? Please give one example for each.
    \item If money wasn't an issue, would you still work as a game companion? Why or why not?
    \item What's the biggest difference you've faced in this job, and how did you solve it?
    \item How has working as a game companion changed your personality and social life the most? Please give an example.
    \item What do your family and friends think of this job? How does it affect your real-life relationships?
    \item Would you take game companion work as a long-term career? What's the main thing you consider when thinking about this?
    \item Why did you stop working as a game companion?
    \item Who is your most memorable client, or what's your most memorable experience in this job?
\end{enumerate}

\end{document}